\documentclass[prb,reprint, amsmath,amssymb,showpacs,superscriptaddress,floatfix]{revtex4-2}
\usepackage[breaklinks=true,colorlinks,citecolor=blue,linkcolor=blue,urlcolor=blue]{hyperref}
\usepackage[table,xcdraw]{xcolor}

\usepackage{epsfig,mathrsfs,color,latexsym,subfigure,marginnote,gensymb,}
\usepackage{graphicx}
\usepackage{xcolor}
\usepackage{booktabs}
\usepackage{appendix}
\renewcommand{\BibitemShut}[1]{}

\begin{document}
\title{Harnessing Room-Temperature Ferroelectricity in Metal Oxide Monolayers for Advanced Logic Devices}

 \author{Ateeb Naseer}
	\affiliation{Department of Electrical Engineering, Indian Institute of Technology, Kanpur, Kanpur 208016, India}
\author{Musaib Rafiq}
\affiliation{Department of Electrical Engineering, Indian Institute of Technology, Kanpur, Kanpur 208016, India}
\author{Somnath Bhowmick}
\email[Corresponding author: ]{bsomnath@iitk.ac.in}
\affiliation{Department of Materials Science $\&$ Engineering, Indian Institute of Technology, Kanpur, Kanpur 208016, India}
\author{Amit Agarwal}
\affiliation{Department of Physics, Indian Institute of Technology, Kanpur, Kanpur 208016, India}
\author{Yogesh Singh Chauhan}
\affiliation{Department of Electrical Engineering, Indian Institute of Technology, Kanpur, Kanpur 208016, India}	
	
\date{\today}
\begin{abstract}
Two-dimensional ferroelectric materials are beneficial for power-efficient memory devices and transistor applications. Here, we predict out-of-plane ferroelectricity in a new family of buckled metal oxide (MO; M: Ge, Sn, Pb) monolayers with significant spontaneous polarization. Additionally, these monolayers have a narrow valence band, which is energetically separated from the rest of the low-lying valence bands. Such a unique band structure limits the long thermal tail of the hot carriers, mitigating subthreshold thermionic leakage and allowing \textcolor{black}{field-effect transistors (FETs)} to function beyond the bounds imposed on conventional FETs by thermodynamics. Our quantum transport simulations reveal that the FETs based on these MO monolayers exhibit a large ON/OFF ratio with an average subthreshold swing of less than 60 mV/decade at room temperature, even for short gate lengths. Our work motivates further exploration of the MO monolayers for developing advanced, high-performance memory and logic devices.   
\end{abstract}
\maketitle

\section{Introduction}
Two-dimensional (2D) materials hold great promise for next-generation electronic, optoelectronic, spintronic, catalysis, energy storage, biomedical, sensing, and other innovative applications~\cite{Pham_app,2D_3, Das2021_2D, Nandan2023, Ateeb_1, Yadav_TMDs, Yoon, Kou, Rodin, Keshari_1, Robin_app, Jin_app, Ateeb2}. These materials are especially appealing due to their tunable electronic structure, high surface-to-volume ratio, dangling bond-free interfaces, enhanced mobility,  and excellent mechanical flexibility~\cite{2D_1,2D_2,2D_3,Nov, Nov1, Nov2}. Therefore, discovering and studying new 2D materials has become an exciting research frontier with unique challenges and opportunities. 

\begin{figure*}[!t] 
	\begin{minipage}[t]{\textwidth}
		\includegraphics[width=1\linewidth]{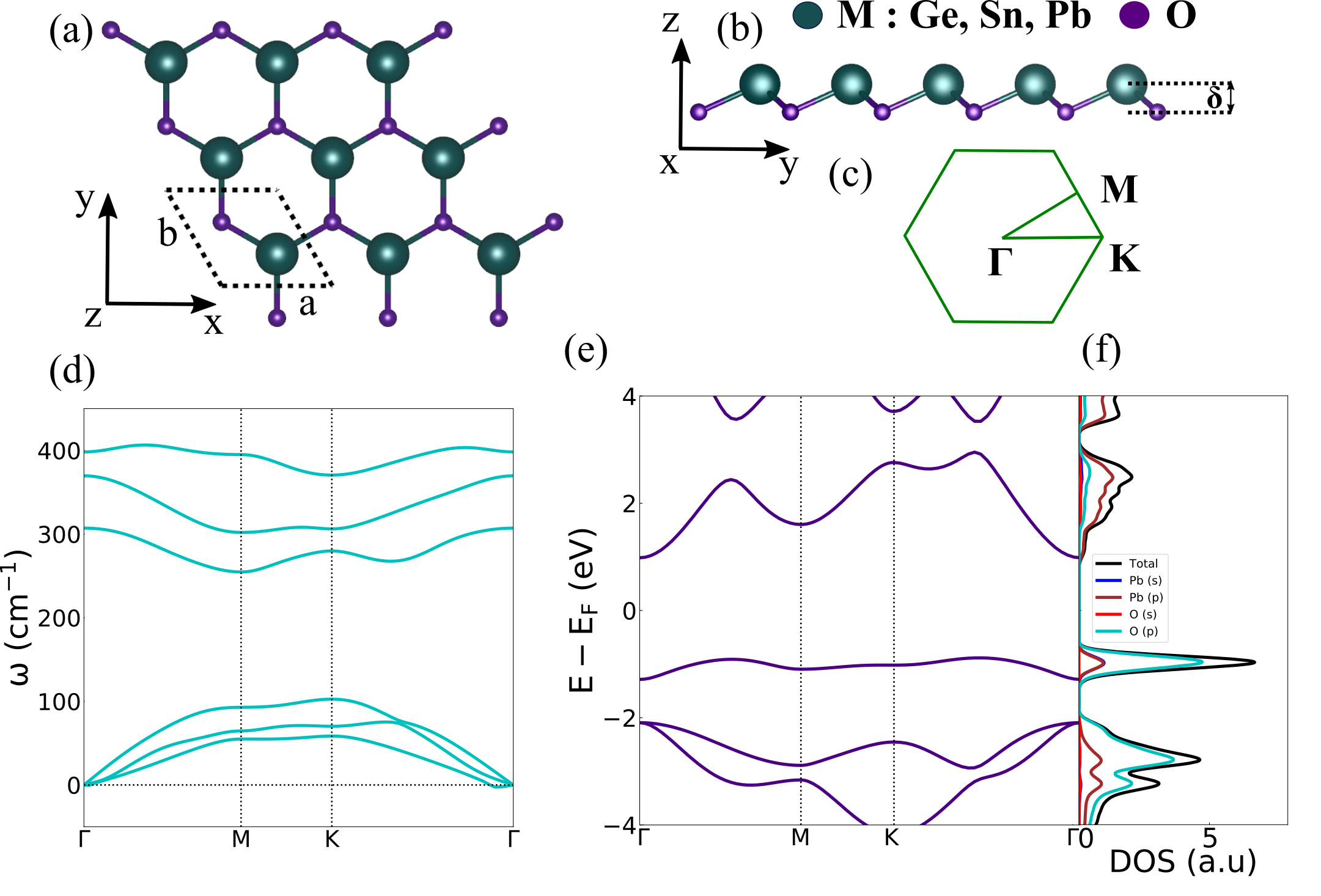}
		\caption{(a) Top view and (b) side view of MO (M: Ge, Sn, and Pb) monolayer crystal with primitive unit cell shown in a dotted enclosure. Buckling height is denoted by $\mathrm{\delta}$. (c) Hexagonal Brillouin zone with high symmetry points. (d) Phonon dispersion curve and (e) electronic bandstructure of monolayer PbO. The absence of negative frequencies in the phonon dispersion curve indicates the dynamic stability of the monolayer PbO. The monolayer PbO is an indirect bandgap material with the valence band maximum between the $\Gamma$ and $\mathrm{M}$ points and the conduction band minimum at the $\Gamma$ point.}
    \label{Fig1}
	\end{minipage}
\end{figure*}

Metal oxides stand out among 2D materials for their unique attributes, such as favorable electronic structures, natural abundance, and chemical inertness~\cite{MO_1, MO_2, MO_3, MO_4, MO_5}. Here, we use first-principle-based methods to investigate monolayer metal oxides (MO; M = Ge, Sn, Pb) and a few related group IV-VI materials. We demonstrate the existence of room-temperature ferroelectricity in these monolayers. The MO-based monolayers remain largely unexplored, with only a few studies on them~\cite{MO_6, MO_7}. However, SnO-based thin films were successfully synthesized recently~\cite{SnO_1}, motivating us to study these MO monolayers systematically. Recent advancements in fabrication methodologies pave the way for fabricating \textcolor{black}{field-effect transistors (FETs)} based on metal oxides with relatively good quality \cite{Fab1, Fab2, Fab3, Fab4}.

We show that all the studied MO monolayers have an out-of-plane spontaneous polarization, the direction of which can be reversed by applying a vertical electric field. The switchable electric polarizations of these materials make them excellent candidates for realizing non-volatile random-access memories~\cite{DEEP_1, Musaib_FE, fei_ferroelectric_2018, Achintya, Ateeb_Nano, PNAS, Musaib_FE1, Robin_app, Piyush, Swetaki_1, Soliman2023, RW_2, Jin_app}. Among these MO monolayers, PbO has the best ferroelectric properties. It has a polarization value of around $\pm$37.6 pC/m and a $\sim$ 162 meV switching barrier. The switchable dual polarization states of the MO monolayers can be used to store data in ferroelectric-based memory devices.  

Additionally, these MO monolayers have a unique electronic band structure, which makes them excellent candidates for ultra-low-power FETs. Conventional  FETs are limited by the thermal leakage current in the off-state from the hot carriers with a long thermal energy tail. This `Boltzmann tyranny' limits the subthreshold swing (SS) to $\geq$ 60 mV/decade at room temperature~\cite{CS_2, Keshari_CS_1, CS_3,CS_4}. In the MO monolayers, the uppermost valence band has a narrow bandwidth and is energetically separated from the remaining low-lying valence bands. Such a band structure restricts the long thermal tail of hot carriers and helps MO-based FETs overcome the limitations of SS $\geq$ 60 mV/decade. Our quantum transport simulations show that the MO FETs have a large ON/OFF ratio with an average SS close to 61 mV/decade for Ge and Sn-based monolayers and an average SS close to 45 mV/decade for Pb-based monolayers with gate length $\sim$ 5 nm. The steep subthreshold characteristics and immunity to source-to-drain tunneling reveal that MO-based FETs are promising candidates for energy-efficient and ultra-low power logic devices. 

\begin{figure*}[!t] 
	\begin{minipage}[t]{\textwidth}
		\includegraphics[width=0.9\linewidth]{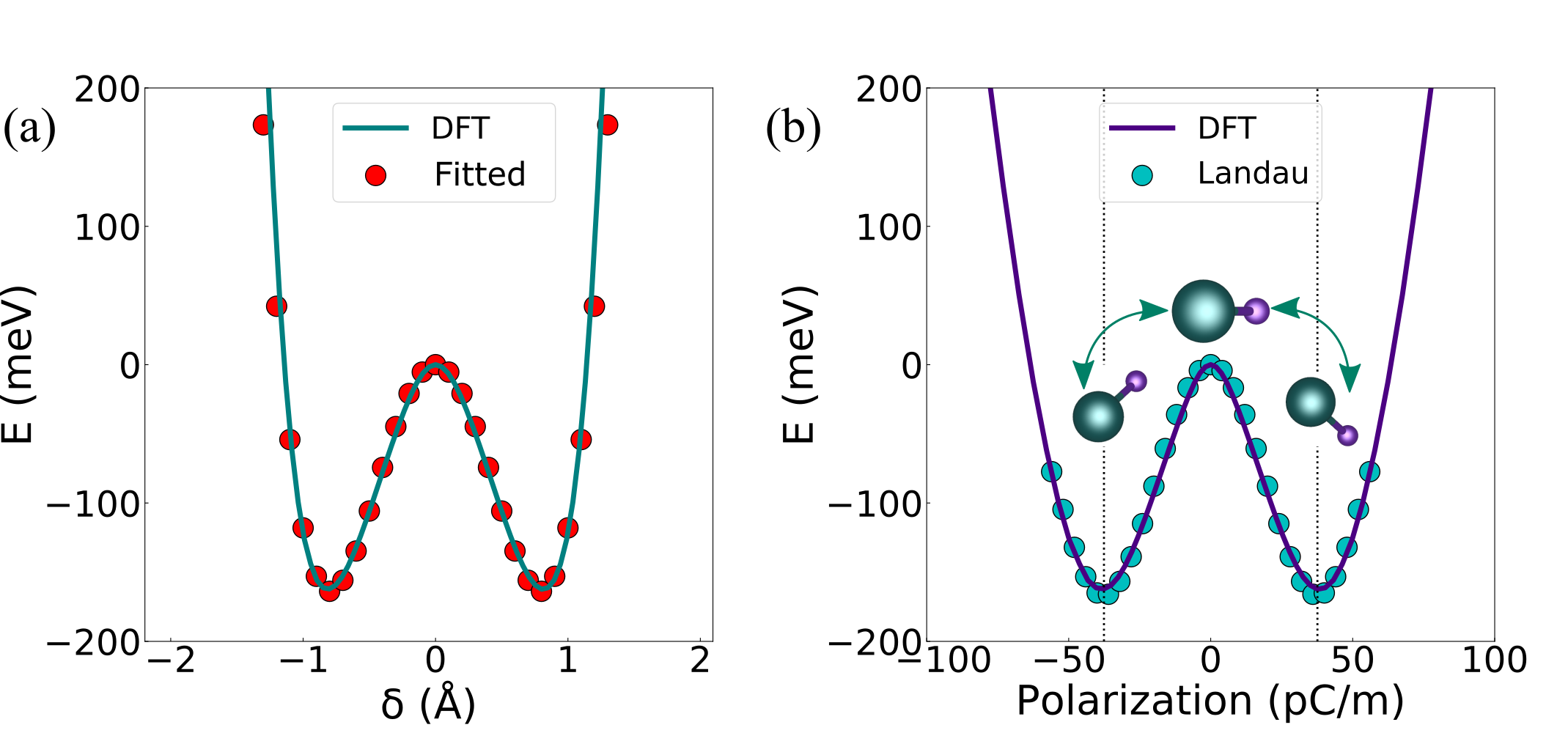}
		\caption{(a) The variation of the free energy with the buckling height ($\delta$) and (b) variation of the free energy with the polarization for monolayer PbO. The transition from one non-centrosymmetric ferroelectric phase to the other via a centrosymmetric paraelectric state with zero polarization is depicted in (b). }
    \label{Fig2}
	\end{minipage}
\end{figure*}

\begin{table*} \label{table1}
		\caption{The potential barrier $E_{G}$ (meV), spontaneous polarization $P_{s}$ (pC/m), Curie temperature ($T_{C}$) in $K$, the coercive field, $E_{c}$ (V/nm), LG fitting parameters in Eq.(1), $A$, $B$ (10$^{-2}$), $C$ (10$^{-5}$), which describe the double-well potential of the energy vs. polarization plots, and $D$, describing the average dipole-dipole interaction.}
		\centering
		\begin{tabular}{||@{}p{0.1\linewidth}|p{0.1\linewidth}|p{0.1\linewidth}|p{0.1\linewidth}|p{0.1\linewidth}|p{0.1\linewidth}|p{0.1\linewidth}|p{0.1\linewidth}|p{0.1\linewidth}@{}||}
			\toprule
            
            \hspace{0.5cm} MO   & \hspace{0.5cm} $E_{G}$ & \hspace{0.5cm} $P_{s}$ & \hspace{0.5cm} $T_C$ & \hspace{0.5cm} $E_{c}$ & \hspace{0.5cm} $A$ & \hspace{0.5cm} $B$ & \hspace{0.5cm} $C$ & \hspace{0.5cm} $D$ \\
            \midrule
            \hspace{0.5cm} GeO &  \hspace{0.5cm} 792.94  &  \hspace{0.5cm} 22.37 & \hspace{0.5cm} 1341   &  \hspace{0.5cm} 7.7    & \hspace{0.5cm} -8.17 & \hspace{0.5cm} 3.287 & \hspace{0.5cm} -6.589 & \hspace{0.5cm} 0.231   \\
            \hspace{0.5cm} SnO & \hspace{0.5cm} 545.01  & \hspace{0.5cm} 23,54 & \hspace{0.5cm} 1900   & \hspace{0.5cm} 4.7  & \hspace{0.5cm} -4.53 & \hspace{0.5cm} 1.492 &\hspace{0.5cm} -2.495 & \hspace{0.5cm} 0.295      \\
            \hspace{0.5cm} PbO &  \hspace{0.5cm} 162.29 &  \hspace{0.5cm} 37.58 & \hspace{0.5cm} 6211   & \hspace{0.5cm} 0.7    \hspace{0.5cm} & \hspace{0.5cm} -0.27 & \hspace{0.5cm} 0.013 & \hspace{0.5cm} -0.00168 & \hspace{0.5cm} 0.379      \\
			\bottomrule
		\end{tabular}
\end{table*}

\section{Results and Discussions}
\subsection{Crystal structure, stability and electronic properties}
Metal oxide monolayers  [general formula MO;  M = Ge, Sn, Pb] have a buckled honeycomb crystal structure belonging to $\mathrm{P3m1}$ space group. We present the top and side view of the crystal structure in Fig.~\ref{Fig1} (a) and (b). Our optimized lattice constants and buckling height $\delta$ agree with values reported previously~\cite{MO_6,MO_7}. This validates our choice of calculation parameters (see Table S1). As the primitive unit cell has two atoms, the phonon dispersion spectrum has three acoustic and optical branches each. The absence of negative frequencies in the phonon dispersion curves confirms their dynamical stability [Fig.~\ref{Fig1} (d) and Fig. S1 (a)-(c) of Supplementary Information (SI)]. The computed electronic band structure of these metal oxides indicates their insulating nature with an indirect bandgap greater than 1.65 eV [see Fig.~\ref{Fig1} (e) and Fig. S1 (d)-(f)]. 

We present the electronic bandstructure of the PbO monolayer along the high symmetry path of $\mathrm{\Gamma}$-M-K-$\mathrm{\Gamma}$ in Fig.~\ref{Fig1} (e). We find that the uppermost valence and lowermost conduction bands are isolated from other bands. This type of bandstructure is well-suited for steep slope FETs as the lack of states restricts the thermionic leakage current~\cite{CS_2, Keshari_CS_1, CS_3}. Our orbital projected density of states (PDOS) calculations show that the p-orbitals of the M atom and p-orbitals of the O atom primarily contribute to the conduction band minima (CBM) and valence band maxima (VBM) as shown in Fig.~\ref{Fig1} (f). 
 
\begin{figure*}[!t] 
	\begin{minipage}[t]{\textwidth}
		\includegraphics[width=0.9\linewidth]{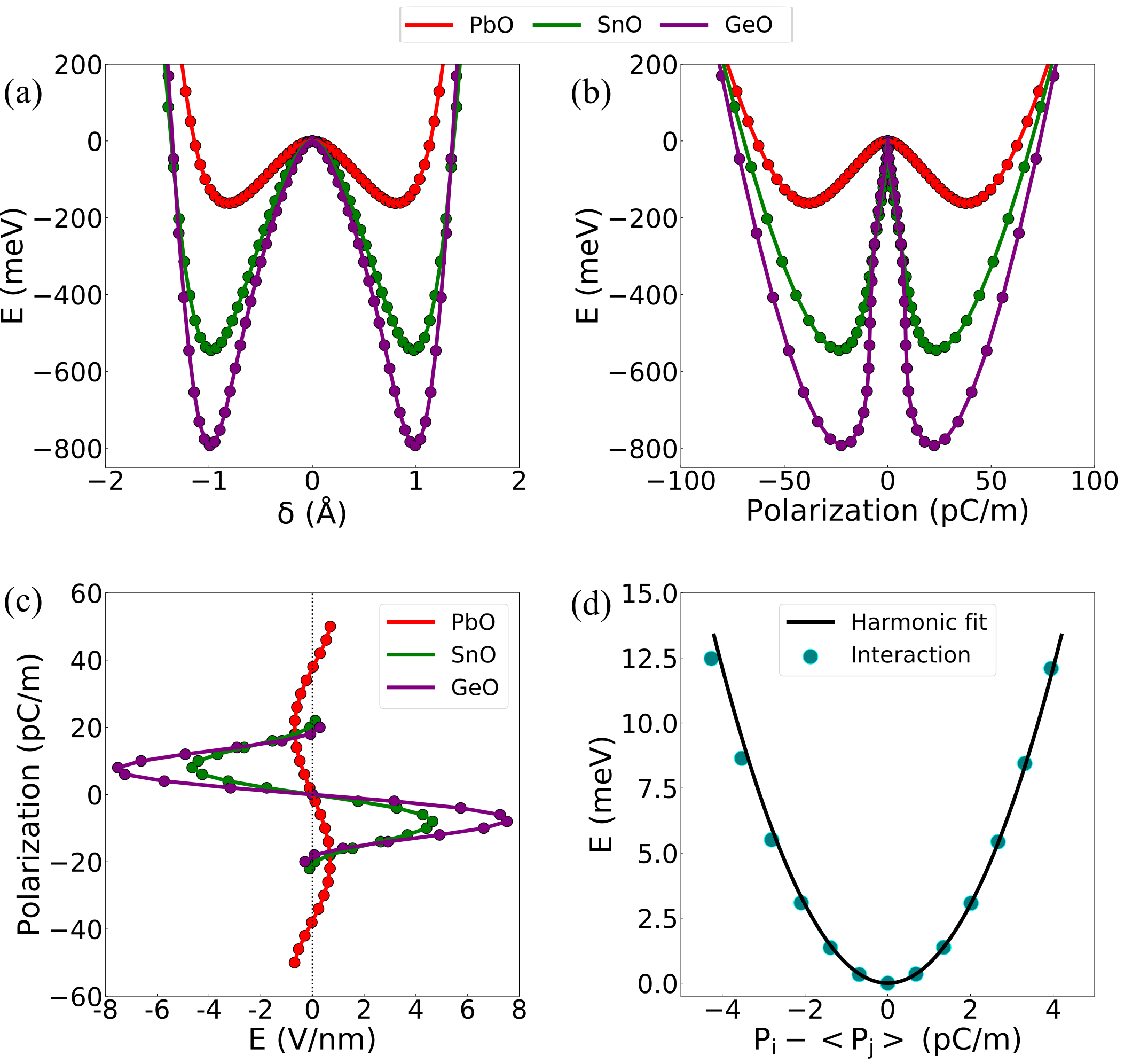}
		\caption{ Variation of the free energy with (a) buckling height ($\mathrm{\delta}$) and (b) polarization, for the Mo series monolayers. (c) The S-curves obtained from the Landau-Ginzburg equation for all the monolayers. (d) The dipole-dipole interaction of monolayer PbO as determined by mean-field theory, with DFT values and harmonic fit represented by circles and a solid line.} 
		\label{Fig3}
	\end{minipage}
\end{figure*}

\begin{figure*}[!t] 
	\begin{minipage}[t]{\textwidth}
		\includegraphics[width=1\linewidth]{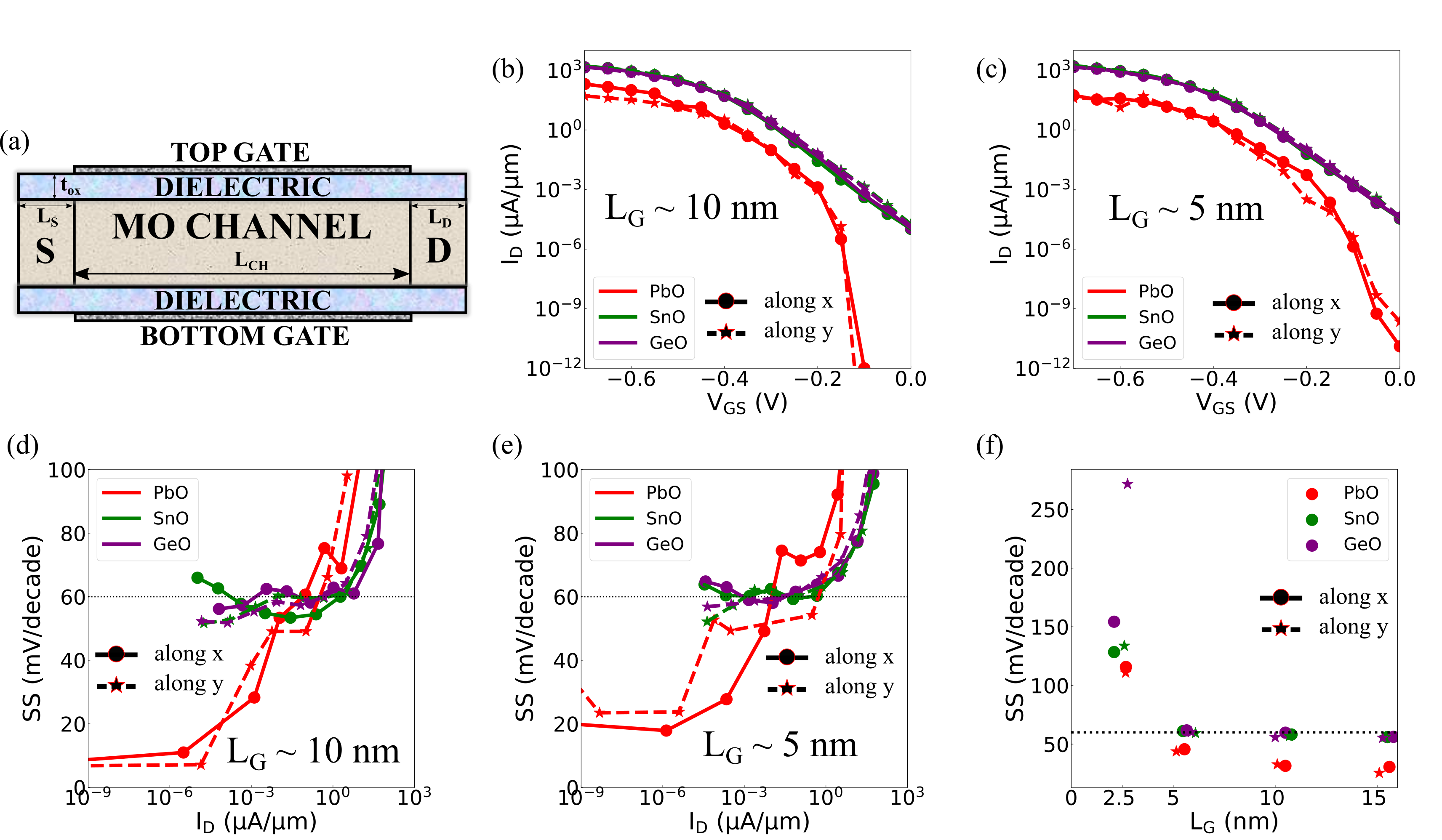}
		\caption{(a) Schematic depiction of double-gate (DG) FETs based on monolayer MO as channel material. $L\mathrm{_{G}}$ ($=$ $L\mathrm{_{CH}}$), $L\mathrm{_{S}}$, $L\mathrm{_{D}}$, and $t\mathrm{_{ox}}$ denote the gate length ($=$ channel length), source length, drain length, and oxide thickness, respectively. Transfer characteristics of each of the three monolayers for $L\mathrm{_{G}}$ $\sim$ (b) 10 nm and (c) 5 nm, respectively, along two transport directions. subthreshold swing vs. drain current of monolayers for $L\mathrm{_{G}}$ $\sim$ (d) 10 nm and (e) 5 nm, respectively, along two transport directions. (f) The average subthreshold swing for four gate lengths along two transport directions. We denote the characteristics with a circle, solid lines along the x-direction, and star, dashed lines along the y-direction. The best-performing FETs are based on monolayer PbO with sub-thermionic characteristics for more than six orders of drain current for gate lengths down to $\sim$ 5 nm. }
    \label{Fig4}
	\end{minipage}
\end{figure*}

\subsection{Ferroelectricity} 
The monolayers based on metal oxides exhibit significant spontaneous polarization in their free-standing forms, confirmed by the first-principle calculations using the Berry phase method~\cite{Berry_1,Berry_3,Berry_2}. Fig.~\ref{Fig2} (a) shows the free energy variation with the buckling height obtained by incremental variation of $\delta$ for the PbO monolayer. Fig.~\ref{Fig2} (b) shows the free energy variation with the polarization and related structural changes leading to the ferroelectric switching for the PbO monolayer. The potential barrier for the ferroelectric switching ($E_{G}$) ranges from 162.29 meV (PbO) to 792.92 meV (GeO) [see Table~\ref{table1}]. It is higher than the potential barrier in elemental ferroelectric like As (5.83 meV) but lower than some other well-known 2D ferroelectric like $\alpha$-In$_2$Se$_3$ (1090 meV)~\cite{Val7,Curie_2}. The stable ferroelectric states are degenerate in energy with opposite spontaneous polarization values ($P_S$). By applying a suitable out-of-plane electric field, one can physically realize the transition from one buckled, non-centrosymmetric, ferroelectric configuration to the other via the planar, centrosymmetric, paraelectric structure [Fig.~\ref{Fig2} and Fig. S2].

Free energy as a function of the buckling height and polarization for the metal oxides are compared in Fig.~\ref{Fig3} (a) and (b), respectively. Apart from exhibiting the lowest barrier, the monolayer PbO has the highest spontaneous polarization values (37.58 pC/m) among the metal oxides under investigation, which is more than three times higher than a well-known 2D ferroelectric $\alpha$-In$_2$Se$_3$ (11 pC/m)~\cite{Val7}. The significant polarization values of monolayer PbO motivate us to further investigate the PbX (X = S, Se) based monolayers [Fig. S3]. The polarization values, along with the energy barriers, are tabulated in Table S2 for PbX monolayers. Monolayer PbS has the highest spontaneous polarization value (47.21 pC/m) but a significant switching barrier (620.61 meV). The trade-off between the polarization value and the switching barrier is essential for memory device applications. Memory device-appropriate materials must have a high spontaneous polarization value and a relatively low potential barrier for ferroelectric switching, allowing for high-speed data writing with low power consumption.

\subsection{Curie temperature}
We express the free energy (G) using the Landau Ginzburg (LG) \cite{LG_1} expansion   
\begin{equation} \label{eq:1}
\begin{aligned}
G= \sum_i \left[\frac{A}{2}P_i^2+\frac{B}{4} P_i^4+\frac{C}{6} P_i^6 \right]\\+\frac{D}{2} \sum_{\langle i, j\rangle}\left(P_i-P_j\right)^2-EP.
\end{aligned}
\end{equation} 
The first three terms describe the anharmonic double-well potential, whose shape is captured by the constants A, B, and C [see Fig.~\ref{Fig2}(b) and Fig.~\ref{Fig3}(b)]. The fourth term accounts for the energy cost of having polarization difference between neighboring unit cells with constant D [Fig.~\ref{Fig3} (d)]. A high value of D implies resistance to the thermal fluctuations and high $T_C$~\cite{Curie_1,Curie_2}. We tabulate the fitting constants in Table~\ref{table1}. We obtain a crude estimate of the Curie temperature ($T_C$) of the monolayers from the relation $K_B$$T_C$ $\simeq$ $D$$\times$$P_{s}{^2}$, with $K_B$ being the Boltzmann's constant in $eV$~\cite{Curie_1}. The estimated Curie temperatures for all the monolayers are greater than 1340 K, suggesting that the investigated monolayers are suitable for room-temperature memory devices and other applications. 

\subsection{S-curve}
The derivative of the free energy curve [Fig.~\ref{Fig2} (b) and Fig.~\ref{Fig3} (b)] with polarization P yields the characteristic S-curve of the ferroelectric monolayers. As illustrated in Fig.~\ref{Fig3} (c), the S-curve gives the polarization variation of the metal oxides under the application of an out-of-plane electric field. The turning points of the  S-curve correspond to the minimum electric field to switch and are known as the coercive field ($E_{c}$). Among the investigated metal oxides, PbO exhibits the lowest $E_{c}$ value of 0.7 V/nm, while GeO exhibits a large $E_{c}$ value of 7.7 V/nm. The $E_{c}$ value of PbO is comparable to the most commonly employed bulk ferroelectric HfO$_2$ ($\sim$ 0.1-0.3 V/nm~\cite{Wang_HfO2}). Since PbO has a reasonable $E_c$ value, we further explore the other PbX-based monolayers and find that the coercive field of PbSe (3 V/nm) $>$ PbS (2.4 V/nm) $>$ PbO (0.7 V/nm). Thus, Pb-based monolayers are desirable for memory device applications in terms of $E_{c}$, which is too high for Sn and Ge-based monolayers. Since Pb-based monolayers, particularly PbO, have high $P_s$ and acceptable $E_c$ values, we explore device applications based on PbO monolayers in the following sections.

\subsection{Steep slope Field Effect Transistors for logic device applications}
It is imperative to design FETs capable of operating at low supply voltage along with steep subthreshold characteristics to enhance the device operations per unit of energy. However, the undesirable subthreshold leakage in the FETs is the predominant factor limiting their application for ultralow power devices. In conventional materials, the hot carriers with long thermal energy tails (Boltzmann tail) resulting from Fermi-Dirac carrier thermal excitation at room temperature are the primary impediments to FET power efficiency. Therefore, it is necessary to eliminate the thermal tail to filter the high-energy carriers and reduce thermionic leakage. MO-monolayers (particularly PbO) have a narrow valence band, which is energetically separated from the rest of the low-lying valence bands, limiting the long thermal tail of the hot carriers, mitigating subthreshold thermionic leakage and allowing p-type FETs to function beyond the bounds imposed on conventional FETs by thermodynamics.

The metal oxides studied have unique electronic band structures with VBM separated in energy from the rest of the low-lying valence bands. The separated VBM with narrow bandwidth restricts the long thermal tail of the carriers, thereby reducing subthreshold thermionic leakage. To evaluate the transport properties of the FETs with metal oxide as the channel material, we evaluate the upper limit performance of the double gate FETs depicted in Fig.~\ref{Fig4} (a) using multiscale ballistic quantum transport simulations based on first principles. The schematic shows the geometrical parameters, with the source, drain, and intrinsic channel composed of monolayer MO. We set the power supply voltage to 0.1 V, equivalent oxide thickness (EOT) $=$ 0.6 nm, and assume the device to be periodic in the transverse direction for all our quantum transport simulations.

The transfer characteristics of MO-based p-type FETs for gate lengths of 10 nm and 5 nm along two transport directions (x and y) are plotted in Fig.~\ref{Fig4} (b) and (c), respectively (see Fig. S4). All the monolayer-based FETs meet the OFF state current ($=$ 10$^{-4}$ $\mathrm{\mu}$A/$\mathrm{\mu}$m) required for the low power FET operation for short channel length FETs down to 5 nm. The plots reveal that even for the 5 nm gate length PbO-based FETs, a large ON/OFF ratio ($>$ 3$\times$ 10$^{4}$) is obtained along both transport directions. Figure~\ref{Fig4} (d) and (e) show the variation of the SS with the drain current for two gate lengths (10 nm and 5 nm, respectively). The PbO-based FETs show sub-thermionic characteristics for more than six orders of drain current for gate lengths down to $\sim$ 5 nm. The SnO and GeO have approximately the same electronic band curvature, as is evident from Fig. S1 (d) and (e), SI. Thus, both monolayers show comparable transfer characteristics.

The degradation of the device characteristics with the decreasing gate length is evident from Fig.~\ref{Fig4} (f). The direct source-to-drain (SDT) tunneling limits the steep subthreshold characteristics of the scaled devices. The effect of SDT is evident for $\sim$ 2 nm gate length FETs with all the monolayers showing degraded average SS of $>$ 100 mV/decade along both transport directions. The PbO-based FETs are more immune to SDT than other MO monolayers with an average SS close to 45 mV/decade for 5 nm gate length FETs along both transport directions. For 5 nm gate length FETs, the average SS in the case of GeO and SnO-based FETs is close to 61 mV/decade.  We simulate the monolayer PbO-based FETs for different EOT values (0.6 nm, 1 nm, and 2 nm), and all of them demonstrate steep subthreshold characteristics, with average SS $<$ 45 mV/decade [Fig. S5, SI]. The steep subthreshold characteristics and immunity to SDT reveal that MO-based FETs, especially PbO-based FETs, are desirable for designing energy-efficient, steep slope logic devices.

\section{Conclusion}  
We use first-principles-based DFT calculations to predict ferroelectricity in monolayer MO oxides (M= Ge, Sn, Pb). The electrically switchable polarization states in the MO monolayers are beneficial for energy-efficient ferroelectric memory devices. Additionally, we show that these monolayers have a distinctive valence band structure with narrow bandwidth and isolation from the other bands, making them highly favorable for energy-efficient, low-power logic devices. 
Their narrow band structure and isolation from other bands reduce the long thermal tail of hot carriers, which are known for generating sizeable thermal leakage currents in conventional FETs and limiting their performance. Our quantum transport simulations highlight GeO/SnO and PbO-based FETs, showcasing impressive scalability and sub-threshold swings. \textcolor{black}{Thin films of SnO have been fabricated and they show chemical stability~\cite{SnO_1}. However, the 2D metal oxides under investigation in this work are still in the initial stages of development and will require further experimental investigations to assess their stability in practical working environments. Furthermore, efforts can be made to enhance their robustness through protective coatings, encapsulation techniques, or controlled fabrication environments.} Our findings provide valuable insights for developing advanced devices from MO monolayers, opening doors to creating more advanced, efficient, and high-performing logic and memory devices. 

\section{Supplementary Material} It includes phonon dispersion and electronic bandstructures of metal oxides, material parameters, free energy vs. buckling height and polarization, ferroelectric properties of the PbX (X: O, S, and Se) series, and device transfer characteristics of the monolayers.

\section*{Acknowledgement}
We acknowledge financial support from the Prime Minister's Research Fellowship (PMRF) and Department of Science and Technology (DST), Government of India, under Grant DST/SJF/ETA-02/2017-18. We acknowledge the National Supercomputing Mission (NSM) for providing computing resources of PARAM SANGANAK at IIT Kanpur, which is implemented by C-DAC and supported by the Ministry of Electronics and information technology (MeitY) and Department of Science and Technology (DST), Government of India. We also acknowledge the HPC facility provided by CC, IIT Kanpur.

\appendix
\section{Methodology}
The density functional theory-based first-principle calculations are carried out using Quantum Espresso \cite{Giannozzi_2009, Giannozzi_2017} suite of codes. The exchange and correlation effects are taken into account using the generalized gradient approximation (GGA) scheme based on Perdew-Burke-Ernzerhof (PBE)~\cite{GGA_PBE} implementation. The kinetic-energy cutoff of the wave function and the charge density are set to 60 $\mathrm{Ry}$ and 600 $\mathrm{Ry}$, respectively. A 12$\times$12$\times$1 Monkhorst–Pack $k$-mesh \cite{Monkhorst} grid is used for Brillouin zone integrations. Along the $\mathrm{z}$-direction, a thick vacuum space ($\sim$ 15 \AA) is added to eliminate the interactions between periodic slabs. 

Afterward, using Wannier90, the Bloch states obtained from ab initio calculations are projected onto a significantly reduced set of localized orbitals (maximally localized Wannier functions)~\cite{Souza, Marzari, Lee}. The overlap matrix between adjacent $k$ points and the projection matrix is the input for calculating Wannier functions. Since the highest valence and lowest conduction bands are primarily contributed by the s and p-orbitals, we use s- and p-orbitals to make the Wannier model. The resulting Tight Binding (TB)-like Hamiltonian is passed to the NanoTCAD ViDES~\cite{NanoTcad_Vides} device simulator with the temperature set to 300 K. Non-Equilibrium Greens' function (NEGF) formalism is used for computing the ballistic quantum transport properties of MO-based FETs~\cite{datta_2005}.
Next, the charge, potential, and transmission coefficient corresponding to each bias are obtained. The transmission coefficient is given by:
\begin{equation}
	\begin{aligned}
			T(E, k)= \text { Trace }{[\Gamma_{S}(E, k) G^{R}.}(E, k)\Gamma_{D}(E, k).G^{A}(E, k)].
		\end{aligned}
\end{equation}
Here, $k$ is the wave vector along the channel width. The periodic boundary condition is implemented by uniformly sampling the transverse wavevector with 30 $k$-points. $\mathrm{G^{R}}$ and $\mathrm{G^{A}}$ = $\mathrm{[G^{R}]^{\dagger}}$ are retarded and advance Green functions, respectively. $\mathrm{\Gamma_{S/D}}$ denotes the broadening from source/drain contacts and is given by:
\begin{equation}
	\begin{split}
	 \Gamma_{S / D}=i[\Sigma_{S / D}-\Sigma_{S / D}^{\dagger}].
	\end{split}
\end{equation} 
where $\Sigma_{S / D}$ are the self-energy matrices of the source and drain contacts, respectively.
Finally, the source-to-drain current is computed using the Landauer-Buttiker approach~\cite{Buttike} and is given by:
\begin{equation}
\begin{split}
	I_{\mathrm{D}}=\frac{2 q}{h}\int_{-\infty}^{\infty} \sum_{k} T(E, k)\times[f_{D}(E-\mu_{D}) \\
	-f_{S}(E-\mu_{S})]d E.
\end{split}
\end{equation} 
Here, ${q}$ is the electronic charge, ${h}$ is the Planck's constant, $\mathrm{\mu_{S / D}}$ is the source/drain chemical potential, and ${f(E-\mu_\mathrm{S})}$ and ${f(E-\mu_\mathrm{D})}$ are the Fermi–Dirac distribution functions at source and drain, respectively.

\bibliography{ref}
\end{document}